# Spindle Nodal Chain in Three-Dimensional α′ Boron

Yan Gao[1], Yuee Xie[1], Yuanping Chen[1]*, Jinxing Gu[2], and Zhongfang Chen[2]

[1]*School of Physics and Optoelectronics, Xiangtan University, Xiangtan, 411105, Hunan, China*

[2]*Department of Chemistry, University of Puerto Rico, Rio Piedras Campus, San Juan, Puerto Rico 00931, United States*

**Abstract:** Topological metal/semimetals (TMs) have emerged as a new frontier in the field of quantum materials. A few two-dimensional (2D) boron sheets have been suggested as Dirac materials, however, to date TMs made of three-dimensional (3D) boron structures have not been found. Herein, by means of systematic first principles computations, we discovered that a rather stable 3D boron allotrope, namely 3D-α' boron, is a nodal-chain semimetal. In the momentum space, six nodal lines and rings contact each other and form a novel spindle nodal chain. This 3D-α' boron can be formed by stacking 2D wiggle α' boron sheets, which are also nodal-ring semimetals. In addition, our chemical bond analysis revealed that the topological properties of the 3D and 2D boron structures are related to the $\pi$ bonds between boron atoms, however, the bonding characteristics are different from those in the 2D and 3D carbon structures.

Corresponding author: chenyp@xtu.edu.cn

Topological semimetals/metals (TM) have recently emerged as a new frontier in the field of quantum materials.[1-5] Different from ordinary three-dimensional (3D) semimetals/metals, TMs possess zero-dimensional (0D) nodal points,[6] one-dimensional (1D) nodal lines[7] or two-dimensional (2D) nodal surfaces[8] around the Fermi level. In the TMs with 0D nodal points, as represented by Weyl-point[9,10] and Dirac-point[11,12] semimetals, conduction and valance bands cross at some points on the Fermi level; in the TMs with 1D nodal lines, the crossings of conduction and valence bands form 1D lines,[13,14] which enable various topology line networks in the Brillouin zone (BZ), such as nodal rings,[15,16] knots,[17,18] links[19-21], chains[22-24] and nets[25]; while in the TMs with 2D nodal surfaces, diverse surfaces induced by band crossings are found,[26] such as plane and sphere,[27] and topological characteristics of the BZ have been clearly divided by the surfaces. These topological elements not only create new topological classifications, but also generate various surface states and electron/hole pockets,[28-31] leading to various exotic electronic, optical, magnetic and superconducting transport properties.[32-34]

On the other hand, boron materials have risen to be star materials recently, and great progresses have been achieved on the experimental syntheses.[35-37] Theoretically, many 2D boron sheets have been proposed, such as $\alpha/\alpha'$-[38,39], $\beta$-[40,41], $\gamma$- and $\chi$-boron[42] sheets. What most famous are honeycomb borophene[43], triangular borophene[44], $\beta_{12}$ and $\chi_3$ boron sheets[45], which have been synthesized successfully on silver or aluminum substrates. α′ boron sheet has also received much attention, because it not only is one of the most stable single-layer boron sheets as revealed by theoretical calculations[39] but also can be basic building blocks of boron allotropes analogous to graphene.[35] Beside these 2D sheets[39, 46], boron has rich 3D structures,[47-49] such as α-Ga,[50] γ-$B_{28}$,[51] and α-$B_{12}$[52]. α-Ga boron phase is expected to be a poor metal exhibiting superconductivity on cooling. γ-$B_{28}$ boron is a semiconductor and a high-pressure phase, whose structure resembles a NaCl, with the $B_{12}$ icosahedra and $B_2$ pairs playing the roles of 'anions' and 'cations', respectively. α-$B_{12}$ boron phase is also a semiconductor, formed by the interconnection of the icosahedral $B_{12}$ clusters. It is found that these three 3D boron phases can transit each other under high pressure. The α-$B_{12}$ to γ-$B_{28}$ phase transformations occur at 19 GPa, and γ-$B_{28}$ to α-Ga phase transformations occur at 89 GPa.[51] In all these boron allotropes, their electron deficient atoms tend to form multiple center bonds,[35] and thus their electronic properties are complicated.[53, 54] For example, the electronic properties of carbon sheets are heavily dominated only by $p_z$ orbital,[55] in contrast, in most boron sheets the energy bands around the Fermi level are attributed by all p orbitals[56]. Therefore, some interesting electronic properties, especially topological properties, are harder to be found in boron structures. To date, diverse topological phases, such as Dirac point,[57] Weyl point,[58] nodal line,[59-61] and nexus network,[62] have been found in 2D and 3D carbon materials. In stark contrast, though a few 2D boron sheets were suggested as Dirac materials,[63-66] all the known 3D boron allotropes so far (experimentally available or theoretically predicted) are insulators (semiconductors) or metals,[48, 49, 51] and none of them is a topological material.

Herein, by systematic density functional theory (DFT) computations, we propose a 3D boron structure, named as 3D-α′ boron (Figure 1a), which can be viewed as a stacking structure of wiggle α′ boron sheets (Figure 1c). The new 3D boron allotrope inherits not only the good stability of α′ boron sheets, but also the bonding characteristics and topological properties of wiggle α′ boron sheets. Interestingly, different from most boron structures, the electronic properties of both 3D-α′ boron and the 2D wiggle boron sheet are dominated by π bonds. The wiggle boron sheet is a nodal-ring semimetal, while the 3D-α′ boron has a spindle nodal chain consisting of six nodal lines and rings because of the interactions between the wiggle sheets. In addition, the π bonds in the structures are confirmed by chemical bonding analysis, and the origination of topological properties is clarified by a tight-binding model based on the π bonds.

All the DFT computations were performed by using Vienna ab *initio* simulation program package (VASP).[67] The Perdew-Burke-Ernzerhof (PBE) functional[68] was employed for the exchange-correlation term according to generalized gradient approximation (GGA). The projector augmented wave (PAW) method[69] was used to represent the ion-electron interaction, and the kinetic energy cutoff of 600 eV was adopted. The atomic positions were fully optimized by the conjugate gradient method, the energy and force convergence criteria were set to be $10^{-5}$ eV and $10^{-3}$ eV/Å, respectively. The phonon calculations were carried out using the Phonopy package[70] with the forces calculated by the VASP code. The nodal-line (nodal-ring) search in Brillouin zone (BZ) and the surface states calculations were performed by using the open-source software Wannier_tools[71] based on the symmetrical Wannier tight-binding model constructed by Wannier90 code.[72]

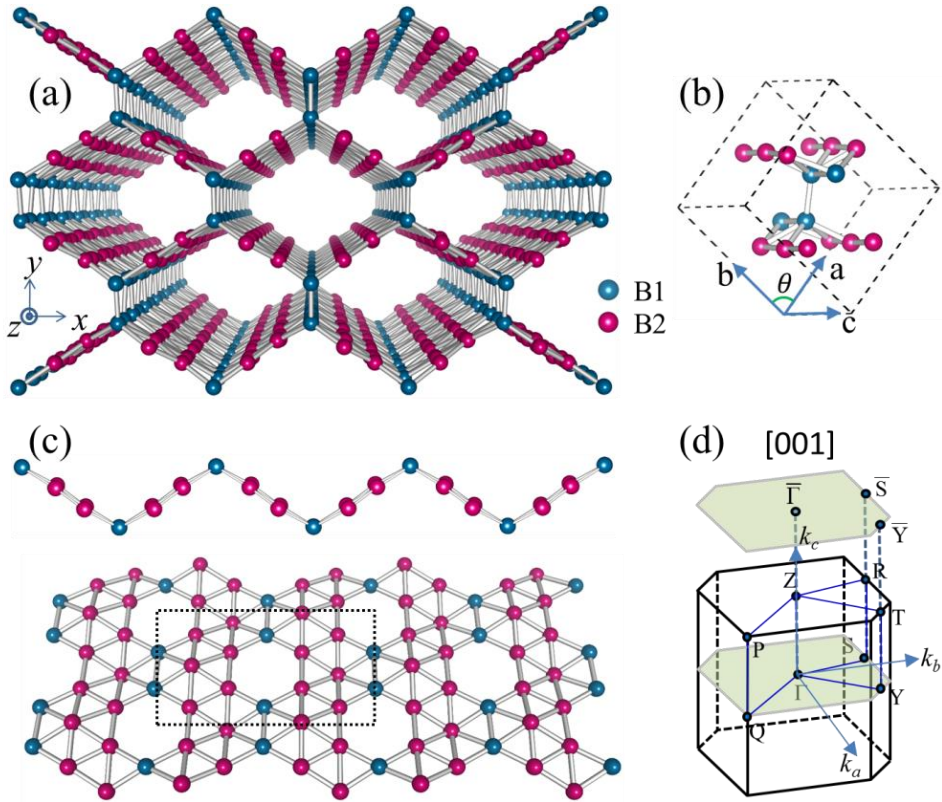

**Figure 1**. (a) Top view of the optimized structure of 3D-α′ boron. (b) The primitive cell of 3D-α′

boron. (c) Side and top views of single-layer wiggle α' sheet and its primitive cell (shown in the dotted box). (d) BZ of 3D-α' boron and projected BZ on [001] surface. For clarity, the atoms linking the single-layer wiggle α' sheet to form 3D-α' boron are labeled as blue B1 atoms, while other atoms are labeled as red B2 atoms.

Figure 1a presents the optimized structure of the 3D-α' boron, which can be constructed by connecting the single-layer wiggle α' sheets via linking atoms (blue atoms labeled as B1 in Figure 1). The wiggle α' sheet is a distorted α' sheet as shown in Figure 1c. As one of the most stable single-layer boron sheets, α' sheet consists of triangular boron lattices and hexagonal vacancies with a hexagonal vacancy density η (defined as the ratio of hexagon holes to the number of atomic sites in the original triangular lattice within one-unit cell)[40] of 1/9 (Figure S1(b)). What differs between α sheet (Figure S1(a)) and α' sheet is that α' sheet is not completely flat, its central atoms in the hexagonal rings are periodically outward or inward. When the wiggle sheets are stacked layer by layer, the blue dimer B1 atoms in the neighbor layers link together, leading to the formation of 3D-α' boron. The primitive unit cell of 3D-α' boron consists of sixteen atoms, as show in the Figure 1b, and these atoms can be divided into two groups: red B2 atoms correspond to the intra-layer atoms in Figure 1a while blue atoms correspond to linking atoms. The symmetry of 3D-α' boron belongs to the space group *CMCM* ($D_{2h}$-17), including two mirror planes, one glide plane, two rotation axes, one screw axis, and one inversion center. The optimized lattice constants are $a = b \cong 5.64$ Å and $c \cong 5.07$ Å, respectively, and the Bravais lattice is a simple monoclinic system with an optimized angle $\theta \cong 86.40°$ between the two lattice vectors in the basal plane. In the optimized structure of 3D-α′ boron, the bond length of dimers in the same α′ sheet is 1.85 Å, and the bonds connecting adjacent α′ sheets are 1.77 Å, while other intralayer boron-boron bonds are in the range of 1.66 to 1.85 Å.

The key structural parameters and properties of 3D-α′ boron is presented in Table 1, the corresponding data for single-layer α′ sheet, α sheet and the three experimentally known 3D boron allotropes are also provided for comparison (the atomic structures of α-$B_{12}$, γ-$B_{28}$, and α-Ga can be seen in Fig. S2). Because of its porous structural feature, 3D-α′ boron has smaller bulk modulus (166.16 GPa) than other 3D boron allotropes (ca. 237~264 GPa), and its density (1.78 g/cm$^3$) is also smaller than others (ca. 2.48~2.81 g/cm$^3$). Analyzing the computed cohesive energies ($E_{coh}$) showed that 3D-α′ boron is a metastable allotrope, since its $E_{coh}$ value (-6.33 eV/atom) is nearly approaching that of α-Ga (-6.41 eV/atom), but is ca. 0.3 eV/atom less favorable than those of γ-$B_{28}$ and α-$B_{12}$.[51] Nevertheless, 3D-α′ boron is dynamically stable, as evidenced by the absence of soft modes in the entire BZ of the computed phonon dispersions (Figure S3)

**Table 1**. Space group, lattice parameters (Å), angles (degrees), Wyckoff position, density (g/cm$^3$), bond lengths (Å), bulk moduli (GPa), and cohesive energy $E_{coh}$ (eV/B) for α-Ga, γ-B$_{28}$, and α-B$_{12}$,[51] 3D-α′ boron, α′ sheet, α sheet.

| Structure | | Space group | Lattice parameters(Å) | | | Angles (degrees) | | Wyckoff position | | | Density (g/cm$^3$) | Bond Lengths(Å) | Bulk Moduli (GPa) | $E_{coh}$ (eV/B) |
|---|---|---|---|---|---|---|---|---|---|---|---|---|---|---|
| | | | a | b | c | α=β | γ | x | y | z | | | | |
| α-Ga | B1(8f) | CMCA | 2.94 | 5.33 | 3.26 | 90 | 90 | 0 | 0.1558 | 0.0899 | 2.81 | 1.76-1.92 | 264.30 | -6.41 |
| γ-B$_{28}$ | B1(4g) | PNNM | 5.04 | 5.61 | 6.92 | 90 | 90 | 0.1702 | 0.5206 | 0 | 2.57 | 1.66-1.90 | 244.29 | -6.65 |
| | B2(8h) | | | | | | | 0.1606 | 0.2810 | 0.3743 | | | | |
| | B3(8h) | | | | | | | 0.3472 | 0.0924 | 0.2093 | | | | |
| | B4(4g) | | | | | | | 0.3520 | 0.2711 | 0 | | | | |
| | B5(4g) | | | | | | | 0.1644 | 0.0080 | 0 | | | | |
| α-B$_{12}$ | B1(18h) | R$\bar{3}$m | 5.05 | 5.05 | 5.05 | 58.04 | 58.04 | 0.0103 | 0.0103 | 0.6540 | 2.48 | 1.67-1.80 | 237.16 | -6.68 |
| | B2(18h) | | | | | | | 0.2211 | 0.2211 | 0.6305 | | | | |
| 3D-α′ boron | B1(16h) | CMCM | 7.73 | 8.23 | 5.07 | 90 | 90 | 0.1696 | -0.2928 | 1.0837 | 1.78 | 1.66-1.85 | 166.16 | -6.33 |
| | B2(8f) | | | | | | | 0 | -0.4008 | 0.9325 | | | | |
| | B3(8g) | | | | | | | 0.1778 | -0.6937 | 1.2500 | | | | |
| α′ sheet | B1(6h) | P$\bar{3}$m1 | 5.06 | 5.06 | 15 | 90 | 120 | 0 | 0.3324 | 0.5 | 0.47 | 1.68-1.70 | --- | -6.28 |
| | B2(2d) | | | | | | | 0.6667 | 0.3333 | 0.5123 | | | | |
| α sheet | B1(6k) | P6/mmm | 5.06 | 5.06 | 15 | 90 | 120 | 0.3311 | 0 | 0.5 | 0.47 | 1.68-1.71 | --- | -6.28 |
| | B1(2d) | | | | | | | 0.6667 | 0.3333 | 0.5 | | | | |

Figure 2a presents the projected band structure of 3D-α′ boron. Clearly, around the Fermi level, crossing points of conduction and valence bands are available on the k paths T-Y, Γ-Y and Γ-S. A careful examination reveals that there are many crossing points in the whole BZ, and they form a nodal line network, as shown in Figures 2(c-e) with different visual angles. The network consists of six nodal lines and rings: two red nodal lines on the plane $k_c = 0$, a green nodal ring on the plane $k_a = k_b$, an orange nodal ring on the plane $k_a = -k_b$, and two blue nodal rings symmetrically on the two sides of the plane $k_a = k_b$. These nodal lines and rings intersect each other and form a spindle nodal chain along the direction $k_a = k_b$ and $k_c = 0$. The space group of 3D-α′ boron is *CMCM*, and its topological elements are protected by different symmetries: the red lines and green ring are protected by the mirror planes $k_c = 0$ and $k_a = k_b$, the orange ring is protected by glide plane $k_a = -k_b$, while the two blue rings are only protected by PT (parity-time) symmetry.[73]

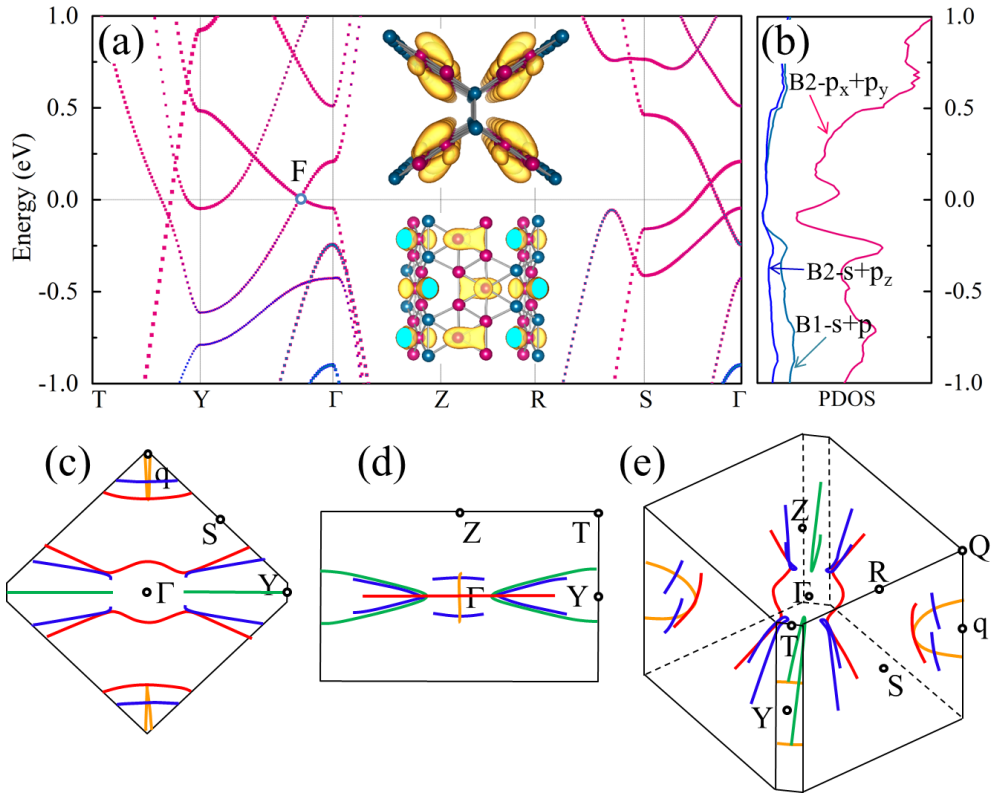

**Figure 2**. Band structure of 3D-α′ boron and its topological phase. (a) Projected band structure of 3D-α' boron, where the pink, blue and cyan bands correspond to $p_x/p_y$ orbitals of atoms B2, s/$p_z$ orbitals of B2 and s/p orbitals of B1, respectively. Inset: the top view and side view of charge density of one state around the crossing point F in (a). (b) The PDOS for 3D-α′ boron, in which the states around the Fermi level are attributed by $p_x$ and $p_y$ orbitals of the atoms B2. (c-e) Top,

side and perspective views of the topological phase in the first BZ, respectively.

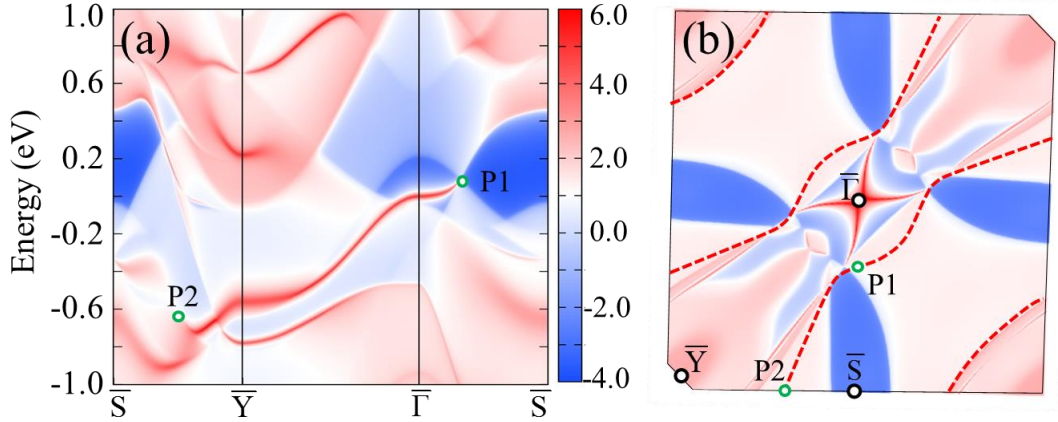

**Figure 3.** Surface states of 3D-α' boron on the [001] surface. (a) Surface energy bands. (b) Surface spectrum cut on a fix energy $E = 0$, where the red dashed lines represent the projection of the red nodal lines on the plane $k_c = 0$ in Fig. 2(c). P1 and P2 are crossing points of bands locate on the projected nodal lines.

The spindle nodal chain is a complicated topological phase, in which the surface states induced by different topological elements (nodal lines or rings) superpose each other and are rather complex. For example, on the surfaces [100] and [110], there exist various surface energy bands related to different nodal lines/rings (Figure S4). Relatively, the energy bands on the [001] surface is clear (Figure 3), because only the two red lines on plane $k_c = 0$ can lead to surface states. Figure 3(a) exhibits the surface energy band in the energy range [-0.8, 0.2] eV. Along $\bar{\Gamma}$ - $\bar{S}$, there is a surface band appearing around the Fermi level. The band starts at P1 and crosses over $\bar{\Gamma}$, extends along $\bar{\Gamma}$ - $\bar{Y}$, and then terminates at the point P2 on $\bar{Y}$ - $\bar{S}$. Figure 3(b) shows the surface spectrum cut on a fix energy *E = 0*, and projections of the two red nodal lines on the plane $k_c = 0$ are also shown as dotted lines. As shown in the surface band in Fig. 3(a), the drumhead states exist between the two nodal lines. The two Fermi arcs around $\bar{\Gamma}$ in Figure 3(b) further prove that there are surface states between the lines.

To explore the origination of the electronic properties of 3D-α′ boron, we calculated its partial density of states (PDOS). As shown in Figure 2(b), the $p_x$ and $p_y$ orbitals of the atoms B2 dominate the energy bands around the Fermi level, while the contributions of other orbitals are insignificant. Analyzing the charge density of one state around the crossing point F revealed that the $p_x$ and $p_y$ orbitals of atoms B2 form π bands rather than σ bonds (see the inset in Figure 2(a)). Note that the π bands here are different from those in graphene: the π bands here link one B atom and a center of another two B atoms; in contrast, in graphene, the π bands directly link two carbon atoms. This difference indicates that the electron-deficient feature of boron and the varied bond lengths may endow the 3D-α′ boron rather complicated bonding patterns.

To understand the bonding features, we performed chemical bonding analysis by utilizing the recently developed solid state Adaptive Natural Density Partitioning (SSAdNDP) method,[74-76] which allows the interpretation of chemical bonding in terms of classical lone pairs and two-center bonds, as well as multi-center delocalized bonding. We chose a unit cell containing 16 B atoms for analysis, in which there are totally 48 electrons or 24 bonds (electron pairs) (Figure 4). According to our analysis, it has two B-B 2c-2e σ bonds between the wiggle α′-sheet with occupation numbers (ONs) ONs = 1.45 |e|, twelve 3c-2e σ bonds around the hexagonal hole with ONs = 1.54/1.57 |e|. The remaining boron skeleton is filled by six 4c-2e σ bonds with ONs = 1.37/1.41 |e|. Especially, there are four 5c-2e π bonds with ONs = 1.53/1.32 |e| in the unit cell, and these four delocalized π bonds are related to the computed charge density of state around the crossing point as presented in Figure 2a. Thus, multi-center bonds are very important to the stability of the 3D-α′ boron, and the 5c-2e π bonds are the key to understand the spindle nodal chain in 3D-α′ boron.

(a) 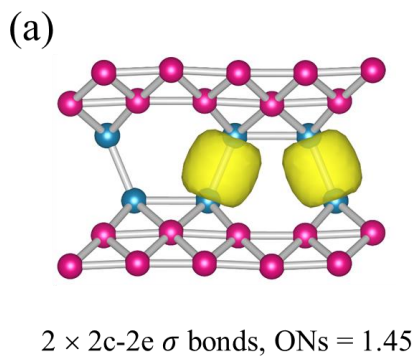

2 × 2c-2e σ bonds, ONs = 1.45

(b) 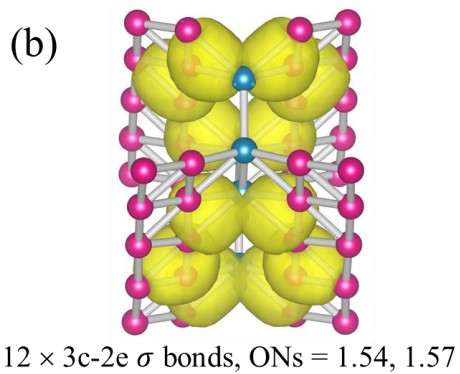

12 × 3c-2e σ bonds, ONs = 1.54, 1.57

(c) 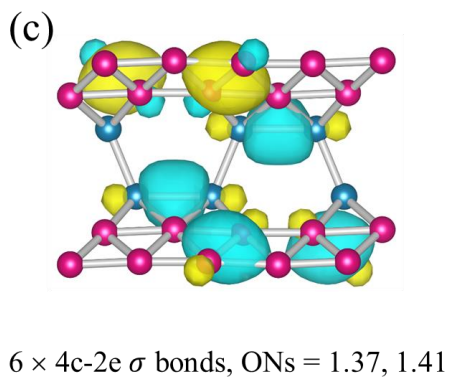

6 × 4c-2e σ bonds, ONs = 1.37, 1.41

(d) 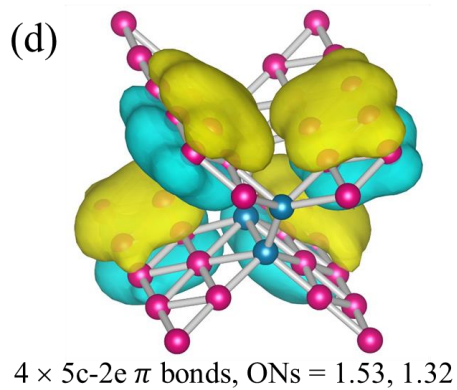

4 × 5c-2e π bonds, ONs = 1.53, 1.32

**Figure 4**. The SSAdNDP chemical bonding analysis of 3D-α′ boron. Totally 24 bonds are exhibited, including (a) two 2c-2e σ bonds, (b) twelve 3c-2e σ bonds, (c) six 4c-2e σ bonds and (d) four 5c-2e π bonds.

Because 3D-α′ boron consists of wiggle α′ born sheets, the topological properties of 3D-α′ boron are also related to the electronic properties of wiggle sheets. Although the bare sheet in Figure 1(c) is unstable, it can stably exist after the B1 atoms are passivated by hydrogen atoms (2D BH-sheet, Figure 5(a)). The space group of 2D BH-sheet is *PMMN*, which is nonsymmorphic and has one glide operation ($G_z$: (x, y, z) → (1/2+x, 1/2+y, -z)), two mirror planes x = 0 and y = 0. Figures 5(b) and 5(c) present its band structure and PDOS, respectively. Around the Fermi level are two crossing points: one is slightly above the Fermi level along Γ-X, and the other is slightly below the Fermi level along Γ-Z (Figure 5b). A closer examination revealed that the two crossing points lie on a ring with a center at Γ in the momentum space (the inset of Figure 5b). Therefore, 2D BH-sheet is a 2D topological nodal-ring semimetal, and the nodal ring is protected by the glide plane $k_y$ = 0. By comparing the topological elements and structural symmetries of 3D-α′ boron and its wiggle sheets, we can find that the orange nodal ring in the 3D boron is inherited from the nodal ring in the 2D sheet.

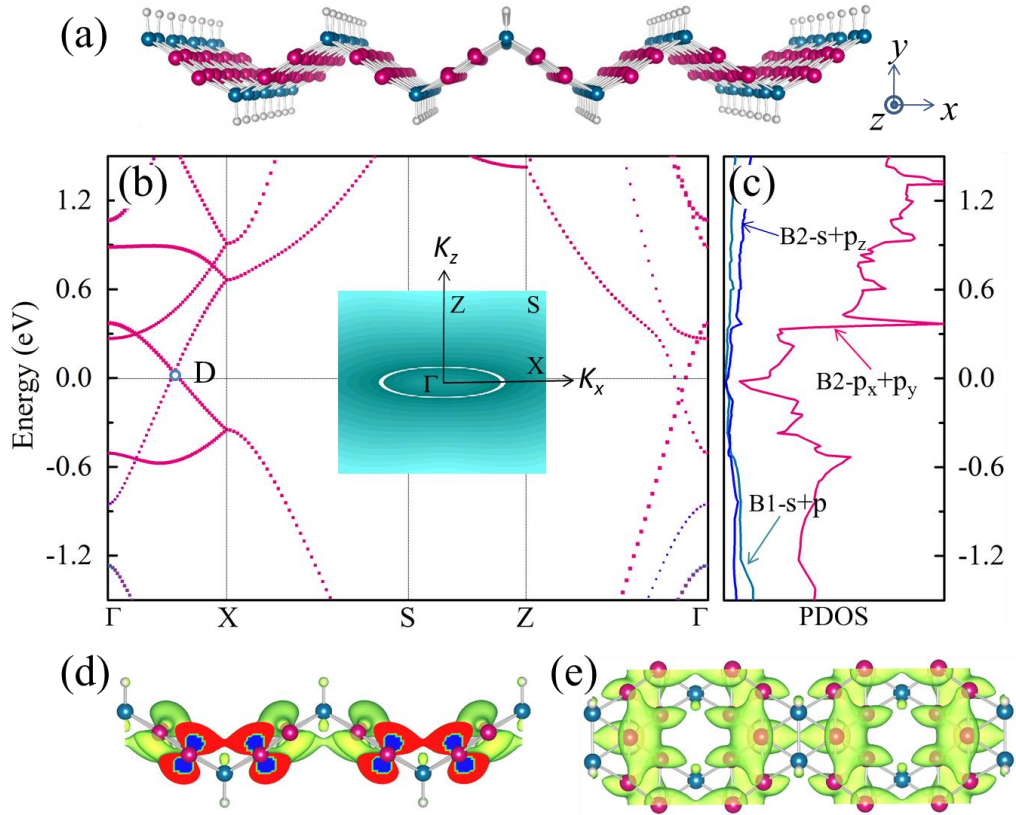

**Figure 5.** Atomic structure and electronic properties of 2D BH-sheet. (a) Optimized geometric structure of 2D BH-sheet whose B1 atoms are passivated by hydrogen atoms. (b) Projected band structure of 2D BH-sheet. Inset: A nodal ring with a center at $\Gamma$ in the momentum space. (c) The PDOS for the 2D BH-sheet, in which the states around the Fermi level are attributed by $p_x$ and $p_y$ orbitals of the atoms B2. (d-e) the side and top views of the charge density of a state around the crossing point D.

The PDOS (Figure 5c) and charge density of a state around the crossing point D (Figures 5d and 5e) illustrate that the electronic properties of 2D BH-sheet is somewhat similar to those of 3D-$\alpha'$ boron: the energy bands around Fermi level are dominated by the $p_x$ and $p_y$ orbitals of the atoms B2, and the $p_x$ and $p_y$ orbitals form $\pi$ bands rather than $\sigma$ bonds. Meanwhile, the 2D BH-sheet has similar bonding pattern (Figure S5) with the 3D-$\alpha'$ boron [Figure 4], as the 3D-$\alpha'$ boron can be constructed by stacking the "layered" 2D sheets and replacing the B-H bonds with 2c-2e B-B bonds.

Because the primitive cell of the wiggle sheet only contains 12 B2 atoms (Figure 1c), and only $\pi$ orbital of each atom needs to be considered, we used a tight-binding (TB) model to describe its electronic properties,

$$H = \sum_{<i,j>} \sum_{\mu} t_{ij} e^{-ik \cdot d_{ij}^{\mu}}, \qquad (1)$$

where $i, j \in \{1, 2, \sim 12\}$ are the 12 B2 atoms in the primitive cell of Figure 1(c), $d_{ij}^{\mu}$ is a vector directed from $j$ to $i$, $t_{ij}$ is the hopping energy between $i$ and $j$, and $\mu$ runs over all lattice sites under translation. The hopping parameters considered here are shown in Fig. S1(c). By tuning the parameters, the band structure based on Eq. (1) can fit the DFT results very well (see Fig. S6(c) in SI). The similar tight-binding model can also successfully simulate the electronic properties of $\alpha/\alpha'$ boron sheets (see Fig. S6). Thus, though the $\pi$ bonds in the $\alpha/\alpha'$ boron sheets and their related structure 2D BH-sheet are 5c-2e, which is completely different from those in graphene, we can use a simple tight-binding model based on a single orbital of each atom to simulate their electronic properties. This provides us an effective tool to study electronic and transport properties of this type of boron materials.

Although the newly predicted metastable 3D-α′ boron has not yet been synthesized, we are rather optimistic about its experimental realization, considering its kinetic stability and the fact that a rich number of metastable boron allotropes, including honeycomb borophene, triangular borophene, $β_{12}$ and $χ_3$ boron sheets, have already been experimentally fabricated. To fabricate the 3D-α′ boron, two possible routes based on the unit of α′ boron sheets (Figure S7) could be considered: (1) Compressing the α′ boron sheets: in this approach, the α′ boron sheets are first stacked to a 3D layered structure, then a compressive strain is applied on the plane of boron sheets. The strong interactions between wiggle sheets could form linkages between layers and thus result in the bulk. Note that this method has been used to prepare the 3D carbon networks from graphene.[77] (2) Dehydrogenation of the stacked 2D BH-sheets: in this process, the 2D BH-sheets are stacked, and the dehydrogenation reaction leads to the formation of 3D-α′ boron.

In conclusion, a 3D boron allotrope, namely 3D-α′ boron, is proposed, and its topological properties are carefully studied. The 3D boron structure consists of 2D wiggle α′ boron sheets. The 2D boron sheet is a nodal-ring semimetal and its electronic properties are attributed by $π$ bonds. 3D-α′ boron inherits the bonding characteristics of the 2D sheet and exhibits interesting topological phase: in the momentum space, six nodal lines and rings intersect each other and form a novel spindle nodal chain.

To the best of our knowledge, this is the first 3D topological boron structure. Because the α/α′ boron sheets can be considered as basic building blocks analogous to graphene to construct 3D structures,[35, 47] one can expect that many 3D boron networks like graphene networks[8, 58, 60-62] can be obtained. These boron networks may exhibit more diverse topological phases because of their rich bonding chemistry. Therefore, our work not only greatly enriches the renown boron family, but also help pave the way to explore exotic topological properties of more 3D boron structures.

## Acknowledgement

We thank Quansheng Wu for useful discussions. This work was financially supported in China by the National Natural Science Foundation of China (No. 11474243 and No. 51376005), and in USA by National Science Foundation-Centers of Research Excellence in Science and Technology (NSF-CREST Center) for Innovation, Research and Education in Environmental Nanotechnology (CIRE2N) (Grant No. HRD-1736093) and NASA (Grant No. 17-EPSCoRProp-0032).